\documentstyle[12pt]{article} %  or [twocolumn,12pt]
\def\a{\alpha}
\def\Acal{\cal A}
\def\Avpl#1{A_{#1, v}^+}
\def\Avperpl#1{A_{#1, v}^{\perp +}} %NB:This def different from that in PI pap.
\def\alphmuoveralphm{\biggl[ \frac{\alpha_s(\mu)}{\alpha_s(m_Q)} \biggr]}
\def\b{\beta}
\def\del{\partial}
\def\Dperp{D_\perp}		\def\Dperpsl{\rlap/\!D_\perp}
\def\Dslash{\rlap/\!D}		\def\Dslashlft{\overleftarrow{\Dslash}}
\def\eq{\begin{equation}}	\def\endeq{\end{equation}}
				\def\endeql#1{\label{#1} \end{equation}}
\def\eqa{\begin{eqnarray}}	\def\endeqa{\end{eqnarray}}
				\def\endeqal#1{\label{#1} \end{eqnarray}}
\def\eqr#1{eq.~(\ref{#1})}	\def\Eqr#1{Eq.~(\ref{#1})}
\def\frac#1#2{{#1 \over #2}}
\def\ga{\gamma^\alpha}
\def\hvpos{h_v^+}
\def\journal#1#2#3#4{{\it #1} {\bf #2}, {#3} (19#4)}
\def\ktilde{\tilde{k}}
\def\Lagr{{\cal L}}
\def\L#1#2{{\cal L}_{{\rm H{#1}EFT},v}^{{\rm gen}#2}}
\def\LmdSI{\Lambda_{\rm SI}}
\def\lvec#1{\overleftarrow{#1}}
\def\oper#1{{\hat O}_{#1}}
\def\ord#1{{\cal O} \left( #1 \right)}
\def\Omga{\Omega^\alpha}
\def\qqquad{\qquad\qquad}
\def\bQv{\bar Q_v}			\def\bQ'v'{\bar{Q'}_{v'}}
\def\smlfrac#1#2{{\textstyle {#1 \over #2}}}
\def\va{v^\alpha}			\def\vpa{v'^\alpha}
\def\vphiv{\varphi_v}
\def\vslash{\rlap/ v}			\def\vtilde{\tilde{v}}
\def\vD{v\cdot D}			\def\vDlft{v\cdot \lvec{D}}
\def\vpD{v'\cdot D}			\def\v'Dlft{v'\cdot \lvec{D}}
\def\vv'{v\cdot v'}

\def\omit#1{}
\newcounter{saveeq}
\def\alpheq{\setcounter{saveeq}{\value{equation}}%
\addtocounter{saveeq}{+1}
\setcounter{equation}{0}%
\renewcommand{\theequation}{\mbox{\arabic{saveeq}\alph{equation}}}}
\def\reseteq{\setcounter{equation}{\value{saveeq}}%
\renewcommand{\theequation}{\arabic{equation}}}

\begin{document}
\begin{titlepage}
\rightline{UCSD-TH-97-24}
\rightline{1997}
\vskip.5in
\begin{center}
 {\Large \bf
  Reparameterization Invariance in Heavy Particle Effective Field Theories}
 \vskip.3in
 {{\bf Clarence L. Y. Lee}\footnote{\tt cl@ucsd.edu}\\
   \vskip.2cm {\it Department of Physics}\\
   \vskip.2cm {\it University of California, San Diego}\\
   \vskip.2cm {\it La Jolla, California 92093-0319}}
 \vskip.2cm
%PACS-96 numbers: 
\end{center}

\vskip.5in
\begin{abstract}

In a low energy effective theory, the fields of heavy particles with mass $m$ 
and total momentum $p$ depend on both a velocity $v$ and a residual momentum 
$k$ such that $p =mv +k$. However, there is some arbitrariness in such a 
description because one may also use a slightly different velocity with a 
compensating change in the residual momentum. This non-uniqueness is the 
origin of an invariance under such a reparameterization which imposes 
non-trivial constraints on the theory. The implications of this invariance 
for effective theories of heavy spin-0, 1/2, and 1 fields is investigated. 

\end{abstract}
\end{titlepage}

%================================ MAIN TEXT ===================================

\section{Introduction}

In recent years, there has been considerable interest in using effective 
field theory methods to analyse hadronic processes involving a heavy 
quark in a model-independent way \cite{IsW89&90,VoS87,Geo90,EiH90,FGL91}.
The utility of this approach, of course, is not limited to heavy quarks. 
Indeed, the low-momentum dynamics of heavy particles of other spins 
interacting with a non-Abelian gauge field may also be conveniently and 
systematically formulated in terms of an effective field theory. When the 
typical interaction scale $\LmdSI$ is much less than the mass $m$ of the 
heavy particle, $\LmdSI \ll m$, the heavy particle is essentially a 
featureless source of colour (or other quantum number) which propagates 
almost on-shell in a nearly straight world-line except for small 
deviations due to its interactions with the gauge field and other light 
degrees of freedom which may be bound to it. Since the interactions of 
such a heavy particle are independent of its quantum numbers (other than 
that it may be in a bound state), this is the situation 
which in the case of QCD is the origin the spin-flavour symmetries 
\cite{IsW89&90,VoS87} which have been so useful in the analysis 
of heavy quark systems. Moreover, in the infinite mass limit the 
velocity $v$ (with $v^2 =1$) of the heavy particle 
is conserved by such soft interactions, and hence, its momentum 
$p$ may be split into a large part which contains the kinematic 
dependence on the heavy mass, $m v$, and a small remaining 
``residual momentum'', $|k^\mu| \sim \LmdSI$ \cite{Geo90}: 
\eq p = m v +k. \endeql{kdef}
A heavy particle effective field theory (HPEFT) can then be constructed 
which incorporates the above picture at leading order in a derivative 
expansion in powers of $k/m$, and the subleading order operators 
correct for the finite mass of the heavy particle. 

The HPEFT is constructed from fields which depend on the velocity of the 
particle \cite{Geo90,Lee97a}. However, when recoil effects are taken into 
account at subleading order in $1/m$, there is no unique assignment of a 
velocity to the heavy particle which leads to an ambiguity in the 
partitioning of the total momentum $p$ \cite{LuM92,DGG92}: 
instead of assigning the particle a velocity and a residual momentum pair, 
$(v, k)$, as in \eqr{kdef}, one may equally well use the pair 
\eq (\vtilde, \ktilde) = (v +\Delta v, k -m \,\Delta v), \endeql{vkVR}
where
\eq p = m \vtilde +\ktilde, \endeq
with $\vtilde^2 =1$ and $|\ktilde| \sim \LmdSI$. The HPEFT expressed 
in terms of these two sets of variables are physically equivalent, 
and this indistinguishability gives rise to a velocity reparameterization 
invariance (VRI) of the theory \cite{LuM92}. It is only the total momentum 
$p$ that is unique which indicates that explicit factors of the velocity 
and residual momentum (or the gauge-covariant derivative in position 
space) should occur in the invariant combination 
\eq \frac{\hat p}{m} = {\hat{\cal V}} = v + \frac{iD}{m}.
\endeql{VRinvart}
While such an observation appears to be trivial, it along with some other 
relations derived below lead to surprisingly powerful constraints on the 
kinds of operators which can appear at subleading order in the $1/m$ 
expansion of the HPEFT. 

Note that although states in the full theory depend on $p$ while those 
in the effective theory carry the two labels $v$ and $k$, extra 
degrees of freedom have not been introduced in the latter because the 
velocity is no longer a dynamical observable but rather a conserved quantity 
\cite{DGG92}.

While there has been an analysis of reparameterization invariance for HPEFT 
before \cite{LuM92}, we have been unable to follow that presentation in its 
entirety, and there is some controversy over the predictions of this 
invariance \cite{FGM97}. In this paper a different approach is used and is 
found to lead 
to some results which appear different from previous ones. In the following 
sections of this paper, the VRI consequences for effective field theories of 
heavy particles with spin 0, 1/2, and 1 are investigated. 

Such an analysis can, in principle, be applied to heavy particles including 
composite ones of any spin, and the predictions are valid to all orders in 
perturbation theory. VRI is also an exact invariance 
of HPEFT --- unlike the spin-flavour symmetries that arise in the heavy 
quark mass limit of QCD, it is not broken by subleading higher-dimensional 
operators which are suppressed by powers of the large mass. Rather such 
operators serve to embody this invariance in the effective field theory. 

We shall begin by considering a scalar field theory before going onto cases 
involving the additional complications of spin and constrained fields.

\section{Reparameterization Invariance of a Heavy Scalar \newline Effective 
Field Theory}

In a previous paper \cite{Lee97a}, the field theory of a heavy scalar 
field $\phi$ of mass $m_s$ with the Lagrangian \cite{Lee97a,LuM92,GeW90}
\eq \Lagr_s = (D_\mu \phi)^\dagger D^\mu \phi - m_s^2\, \phi^\dagger \phi,
\endeql{L_s}
where
\eq D^\mu \phi = (\del^\mu - ig\Acal^\mu)\phi \endeq
was examined. Here, terms in the Lagrangian involving external sources 
have been dropped. In particular, using a functional integral formalism 
an effective field theory (HSEFT) was constructed which reproduces 
the physics of the above full theory Lagrangian, \eqr{L_s}, at scales below 
the large mass $m_s$. In this paper, we re-examine the kind of constraints 
that are placed on this HSEFT by reparameterization invariance. 

As in ref. \cite{LuM92,Lee97a} the effective theory is constructed for scalar 
particles so that it is expressed in terms of the field $\vphiv$ which creates 
and annihilates scalars of a definite velocity and residual momentum 
but does not act on anti-scalars. 
Under a reparameterization, as described above, the velocity and 
residual momentum transform, respectively, as 
\alpheq
\eq v \rightarrow \vtilde = v +\Delta v, \endeql{vVR}
and 
\eq k \rightarrow \ktilde = k -m_s \>\Delta v. \endeql{kVR}
\reseteq
Since the full theory field $\phi_v$ defined by
\alpheq
\eq \phi_v (x) = \phi_v^+(x) +\phi_v^-(x) =e^{im_s v\cdot x} \phi(x), \endeq
transforms by an overall phase factor, and the effective theory field 
$\vphiv =\phi_v^+$ is related to $\phi_v$ by \cite{Lee97a}
\eq \vphiv = \left(1 +{i\vD \over 2m_s}\right) \phi_v, \endeq
$\vphiv$ transforms as 
\alpheq
\eqa \vphiv \rightarrow \tilde{\vphiv} \hskip -.5em
 &=& \hskip -.5em e^{im_s\Delta v \cdot x}  \Biggl\{\vphiv
  +\Biggl[{i\Delta v \cdot D \over 2m_s} +\ord{(\Delta v)^2} \Biggr] \phi_v
   \Biggr\} \label{vphivVRexact}\\
 &=& \hskip -.5em e^{im_s\Delta v \cdot x}
  \Biggl\{ 1 +{i\Delta v \cdot D \over 2m_s}
  \left[1 -{1\over 2m_s (2m_s +i\vD) +(\Dperp)^2} (\Dperp)^2 \right]
  +\ord{(\Delta v)^2} \Biggr\} \vphiv \nonumber\\ \hskip -.5em 
 &=& \hskip -.5em e^{im_s\Delta v \cdot x}
  \Biggl\{1 +{i\Delta v \cdot D \over 2m_s}
  \left[1 -{(\Dperp)^2 \over 4m_s^2} -{2m_s i\vD +(\Dperp)^2 \over 16m_s^4} 
   (\Dperp)^2 \right. \nonumber\\
 && \hskip 4em
  +\left.\ord{(\Delta v)^2, \smlfrac{\Delta v}{m_s^5}}\right]\Biggr\} \vphiv.
   \label{vphivVRapprox}
\endeqa
\reseteq
The relation in \eqr{vphivVRexact} is exact. However, in the two lines 
following it, a tree-level result from the path integral approach in ref. 
\cite{Lee97a} has been used to express the antiscalar component $\phi_v^-$ in 
terms of $\vphiv$. The $\phi_v^-$ component does not contribute to 
the transformation of $\vphiv$ until order $1/m_s^3$ so \eqr{vphivVRapprox} 
will not receive radiative corrections and hence be valid up to at least this 
order. Such loop corrections which may be calculated, for instance by 
matching a three-point function consisting of two scalar and a gauge 
external line, will generally modify the terms in \eqr{vphivVRapprox} 
beyond order $1/m_s^3$. 

Since this effective theory field, even though it represents a 
spinless particle, transforms non-trivially under a velocity 
reparameterization, it is now clear that not only must the velocity and 
covariant derivative appear in the combination in \eqr{VRinvart}, but there 
are also additional constraints on the effective Lagrangian coming from the 
above transformation of the fields, \eqr{vphivVRexact}. 

The constraints imposed by such a reparameterization symmetry in the 
effective theory may be determined by examining its effects on the most 
general set of operators. The most general effective Lagrangian subject 
to the general requirements of gauge, space inversion, and time 
reversal invariance, locality, and hermiticity, 
which corresponds to the full theory Lagrangian above is
\eq \Lagr_{\rm HSEFT}^{\rm gen} = \sum_v \L{S}{},\endeq
where
\eq \L{S}{} = \sum_{j=0}^\infty \frac{\L{S}{(j)}}{(m_s)^{j-1}}, \endeq
and the first several terms are
\alpheq
\eqa
 \L{S}{(0)}&=& 2 \vphiv^\dagger i\vD \vphiv, \\
 \L{S}{(1)}&=& \vphiv^\dagger \left[a_{1a} D^2 +a_{1b} (\vD)^2 \right]\vphiv,\\
 \L{S}{(2)}&=& i\vphiv^\dagger \left[a_{2a} D_\mu \vD D^\mu
      +a_{2b} \left(D^2 \vD + \vD D^2 \right) +a_{2c} (\vD)^3 \right] \vphiv,\\
 \L{S}{(3)}&=& \vphiv^\dagger \left\{a_{3a} D^4
      +a_{3b} \left[D^2 (\vD)^2 + (\vD)^2 D^2 \right] +a_{3c} (\vD)^4 \right.
      \nonumber \\
  &&  +a_{3d} (D_\mu \vD D^\mu \vD + \vD D_\mu \vD D^\mu)
      +a_{3e} D_\mu (\vD)^2 D^\mu \nonumber \\
  &&\left.+a_{3f} D_\mu D^2 D^\mu
      +a_{3g} D_\mu D_\nu D^\mu D^\nu \right\}\vphiv,\\
 \L{S}{(4)}&=& i\vphiv^\dagger 
      \left\{ a_{4a} (D^4 \vD +\vD D^4) +a_{4b} D^2 \vD D^2 \right.\nonumber \\
  &&  +a_{4c} \left(D^2 D^\mu \vD D^\mu +D^\mu \vD D_\mu D^2 \right)
      +a_{4d} \left(D_\mu D^2 \vD D^\mu +D^\mu \vD D^2 D_\mu \right)\nonumber\\
  &&  +a_{4e} D_\mu D_\nu \vD D^\mu D^\nu +a_{4f} D_\mu D_\nu \vD D^\nu D^\mu
  \nonumber \\
  &&  +a_{4g} (D_\mu D_\nu D^\mu \vD D^\nu +D^\nu \vD D^\mu D_\nu D_\mu)
      +a_{4h} \left[D^2 (\vD)^3 +(\vD)^3 D^2 \right] \nonumber\\
  &&  +a_{4i} D_\mu (\vD)^3 D^\mu +a_{4j} \vD D_\mu \vD D^\mu \vD \nonumber \\
  &&  +a_{4k} \left.\left[\vD D^2 (\vD)^2 +(\vD)^2 D^2 \vD \right]
      +a_{4l} (\vD)^5 \right\} \vphiv.
\endeqa
\reseteq
The coefficients denoted by ``$a$'' have been introduced to take into 
account short-distance contributions, and the sum over the different 
velocities is needed to recover Lorentz covariance.

The field may be redefined to eliminate the subleading terms in 
$\L{S}{(j)}$ which vanish by the leading order equation of motion, namely 
$i\vD \vphiv =0$, and would lead to an operator basis considerably simpler 
than the one above. However, the redefined field will then transform under a 
reparameterization not as in eq. (\ref{vphivVRexact}-\ref{vphivVRapprox}) 
but in a more complicated way with extra terms containing $\vD \vphiv$ so 
that the analysis is perhaps even more labourious. Moreover the field 
redefinition must be done at each successive order in $1/m_s$ which is not 
very convenient. Hence in this and the subsequent analyses, operators which 
vanish by the leading order equation of motion are retained. 

This Lagrangian must remain invariant under a velocity reparameterization 
as given by eq.~(\ref{vVR}-\ref{kVR},\ref{vphivVRexact}-\ref{vphivVRapprox}): 
\eq \Delta \L{S}{}
 = \Lagr_{{\rm HSEFT},\vtilde}^{\rm gen} - \L{S}{} = 0. \endeq
Imposing this condition on $\L{S}{}$ at each order in $1/m_s$ up to 
operators of order $1/m_s^2$, which is certainly within the range of 
validity of \eqr{vphivVRapprox}, yields the following constraints 
for the coefficients.
\eqa
 a_{1a} = -1, \nonumber\\
 a_{1b} - a_{2a} - 2a_{2b} = 1, \nonumber\\
 a_{2b} + 2a_{3a} + a_{3f} + a_{3g} = \frac{1}{2}, \nonumber\\
 \frac{a_{1b}}{2} + a_{2c} + 2a_{3b} + a_{3d} + a_{3e} = 0, \nonumber\\
 a_{2a} + 2a_{3f} + 2a_{3g} = 0, \nonumber\\
 a_{2c} + 2a_{3d} = 0, \nonumber\\
 a_{3b} - 2a_{4a} - a_{4c} = -{1 \over 4}, \nonumber\\
 a_{3c} - a_{4j} - 2a_{4k} = {1 \over 4}, \nonumber\\
 -{a_{2a}\over 2} +a_{3d} - 2a_{4c} - a_{4e} - a_{4f} - a_{4g} = 0, \nonumber\\
 -{a_{2b}\over 2} +a_{3b} - 2a_{4b} - a_{4c} - a_{4d} - a_{4g} = 0, \nonumber\\
 {a_{2b} \over 2} + 2a_{4a} + a_{4d} = 0, \nonumber\\
 -{a_{2c} \over 2} + a_{3c} - 2a_{4h} - a_{4i} = 0, \nonumber\\
 a_{3d} = a_{4g}, \nonumber\\
 a_{3e} - 2a_{4d} - a_{4e} - a_{4f} - a_{4g} = 0.
\endeqal{HSEFTRI}

The VRI constraints for this scalar field theory derived here appear to 
differ from those obtained by Luke and Manohar in ref. \cite{LuM92}. There 
they found that the effective theory field transforms by a phase under a 
velocity reparameterization and hence concluded that for the scalar field 
theory Lagrangian to be reparameterization invariant, it was necessary and 
sufficient for the factors of $v$ and $D$ to always occur in the 
combination in \eqr{VRinvart}. Here we find that the field in the 
effective theory transforms under a reparameterization not by just a phase 
but in accordance to eq.~(\ref{vphivVRexact}-\ref{vphivVRapprox}), so that 
their condition seems to be necessary but not sufficient. Since their 
transformation does not have any contributions which are suppressed by powers 
of $m_s$, it now easy to see that while their leading order prediction, 
namely, $a_{1a} =-1$, is in agreement with what is found here, the other 
constraints would be expected to differ. The VRI predictions for the 
operators in $\L{S}{(j)} \hbox{ for } j\geq 2$ were not calculated in 
ref. \cite{LuM92}, and since the interpretation of eq. (2.9) in that paper 
is not completely clear to us ({\it i.e.}, whether the most general 
reparameterization invariant effective Lagrangian can be constructed by 
writing down all possible operators assembled out of their $\phi_v$ and 
${\cal V}$ or otherwise), we are unable to make a direct comparison. 

In ref. \cite{Lee97a}, the tree-level matching between the full and effective 
theory Lagrangians was performed. It is important to compare the results 
from that calculation with the ones here required by VRI. The tree-level 
coefficients were found to be 
\eqa
 -a_{1a} = a_{1b} =1 \nonumber\\
  a_{2a} = a_{2b} = a_{2c} = 0 \nonumber\\
  a_{3a} = -a_{3b} = a_{3c} = \frac{1}{4} \nonumber\\
  a_{3d} = a_{3e} = a_{3f} = a_{3g} = 0 \label{HSEFTPI}\\
 -a_{4b} = a_{4h} = -a_{4l} = \frac{1}{8} \nonumber\\
  a_{4a} = a_{4c} = a_{4d} = a_{4e} = a_{4f} = a_{4g} = a_{4i} = a_{4j}
  = a_{4k} = 0 \nonumber
\endeqa
which are consistent with the VRI constraints obtained above. 

Next we shall apply these considerations to particles with spin.

\section{Reparameterization Invariance of a Heavy Spin-$\smlfrac{1}{2}$ 
\newline Fermion Effective Field Theory}

In this section we formulate a framework for applying a 
reparameterization invariance analysis to heavy spin-$\smlfrac{1}{2}$ 
particles. While this method can be used for any such spin-$\smlfrac{1}{2}$ 
particle, for concreteness we shall focus specifically on the physical system 
where it has been most useful~--- namely to the low-energy strong 
interactions of heavy quarks. 

A natural way to describe the low-momentum QCD 
interactions of heavy quarks is in terms of an effective field theory 
(HQEFT) which is formulated in terms of an expansion in inverse powers 
of the large mass with operators containing velocity-dependent heavy 
quark fields \cite{Geo90}.  Using the notation in ref. \cite{Lee97a}, one 
starts with the QCD Lagrangian for the heavy quark:
\eq \Lagr_{H,\rm QCD} = \bar \psi (i\Dslash -m_Q) \psi. \endeql{L_H-QCD}
The heavy quark effective field $\hvpos$ of HQEFT defined through 
\alpheq
\eq \psi_v'(x) = e^{im_Q v\cdot x} \psi_v(x) \endeq
\eq h_v^\pm(x) = {1 \pm \vslash \over 2} \psi_v'(x) \endeql{hvpmdef}
\reseteq
acts on quarks of velocity $v$ but not on antiquarks. The transformation 
of this field under the above velocity reparameterization 
eq.~(\ref{vVR}-\ref{kVR}) follows from the observation that the field 
$\psi_v'$ transforms by a global phase factor so that the definitions in 
\eqr{hvpmdef} and ref. \cite{Lee97a} give
\alpheq
\eqa \hvpos \rightarrow \tilde{\hvpos}
 &=& e^{im_Q \Delta v \cdot x}  \left[ \hvpos
      +\frac{\gamma_\mu\, (\Delta v)^\mu}{2} h_v \right] \label{hvposVRexact}\\
 &=& e^{im_Q \Delta v \cdot x} \Biggl[1 +\frac{\gamma_\mu\, (\Delta v)^\mu}{2}
   \Biggl(1 +\frac{1}{2m_Q +i\vD}i\Dperpsl \Biggr)\Biggr]\hvpos \\
 &=& e^{im_Q \Delta v \cdot x} \Biggl\{1 +\frac{\gamma_\mu\, (\Delta v)^\mu}{2}
   \Biggl[1 +\frac{i\Dperpsl}{2m_Q} -\frac{i\vD}{2m_Q}\frac{i\Dperpsl}{2m_Q}
    +\left(\frac{i\vD}{2m_Q}\right)^2 \frac{i\Dperpsl}{2m_Q} \Biggl]\nonumber\\
 && \hskip 4em +\ord{(\Delta v)^2, \frac{\Delta v}{m_Q^3}} \Biggr\}\hvpos. 
  \label{hvposVRapprox}
\endeqa
\reseteq
\Eqr{hvposVRexact} is an exact expression but in the last two equalities 
above, $h_v^-$ has been expressed in terms of $\hvpos$ using a relation in 
ref. \cite{Lee97a} that holds to order $1/m_Q$. Once again, radiative 
contributions will likely introduce corrections beyond this order. 

To simplify the notation, 
we shall set
\eq Q_v = \hvpos. \endeql{Qvdef}
Then to investigate the implications of reparameterization invariance on the 
effective Lagrangian, we first determine the most general form of this 
Lagrangian consistent with the QCD symmetries of the theory: 
\eq \Lagr_{\rm HQEFT}^{\rm gen} = \sum_v \L{Q}{},\endeq
with
\eqa \L{Q}{}
 &=& \bQv i\vD Q_v \nonumber\\
 &&\hskip -1em +\frac{1}{2m_Q}\bQv 
     \left[b_{1a}(iD)^2 -b_{1b}(i\vD)^2
      +\frac{b_{1c}}{2}g_s \sigma^{\mu\nu} G_{\mu\nu} \right] Q_v \nonumber\\
 &&\hskip -1em +\frac{1}{4m_Q^2} \bQv
     \left[\frac{b_{2a}}{2} g_s v^\mu [D^\nu,G_{\mu\nu}]
           +\frac{ib_{2b}}{2} g_s\sigma^{\alpha\mu}v^\nu\{D_\alpha,G_{\mu\nu}\}
           +\frac{ib_{2c}}{2} \{D^2,\vD\} \right. \nonumber\\
  &&\hskip 4em \left. +ib_{2d} (\vD)^3
           +\frac{ib_{2e}}{2} g_s \{\sigma^{\mu\nu} G_{\mu\nu},\vD\}\right]Q_v.
\endeqal{LgenHQEFTv}
In anticipation of their future use, we define the following operators. 
\eqa
 \oper{1a} &=& {1\over 2m_Q} \bQv\, (iD)^2 \, Q_v, \nonumber\\
 \oper{1c} &=& {g_s\over 4m_Q} \bQv\, \sigma^{\mu\nu} G_{\mu\nu} \, Q_v,
   \nonumber\\
 \oper{2a} &=& {g_s\over 8m_Q^2} \bQv\, v^\mu [D^\nu,G_{\mu\nu}]\, Q_v,
   \label{LHQEFTOp}\\
 \oper{2b} &=& {ig_s\over 8m_Q^2} \bQv\, \sigma^{\alpha\mu} v^\nu
                \{D_\alpha,G_{\mu\nu}\}\, Q_v. \nonumber
\endeqa

Now consider a small reparameterization of the velocity and residual momentum 
as given by eq.~(\ref{vVR}-\ref{kVR}), then for physical results to remain 
invariant, it is necessary to require that
\eq \Delta \L{Q}{}
 = \Lagr_{{\rm HQEFT},\vtilde}^{\rm gen} -\L{Q}{} = 0. \endeq
For a small but arbitrary shift of $v$ by $\Delta v$ and $k$ by 
$\Delta k = -m_Q\, \Delta v$, this relation requires that up to order $1/m_Q$
\eqa
 b_{1a} &=& 1, \nonumber \\
 b_{2b} &=& 2 b_{1c} -1, \nonumber \\
 b_{2c} &=& 2 b_{1b} -1. \endeqal{RILHQEFTpred}

Since these relations should hold to arbitrary order in the coupling, 
it is essential to verify that they are satisfied by perturbative 
calculations. The coefficients of the $\ord{1/m_Q}$ operators have been 
to calculated to one-loop order in \cite{EiH90,FGL91,1/m^2eg} and are 
found to be 
\eqa
 b_{1a}(\mu) &=& 1 +\ord{\alpha_s(\mu)^2}, \nonumber \\
 b_{1b}(\mu) &=& 3\alphmuoveralphm^{-{8\over 33-2n_f}}-2, \nonumber \\
 b_{1c}(\mu) &=& \alphmuoveralphm^{-{9\over 33-2n_f}},
\endeqa
where $n_f$ is the number of effective light quark flavors in to the 
momentum interval between $\mu$ and $m_Q$. Coefficients of some of the 
$\ord{1/m_Q^2}$ operators have also been calculated to the same order; 
in particular \cite{1/m^2eg}
\eq  b_{2b}(\mu) = 2 \alphmuoveralphm^{-{9 \over 33-2n_f}} -1, \endeq
so these results are consistent with the VRI predictions above, 
\eqr{RILHQEFTpred}, as required. 

The transformation of $Q_v$ under a reparameterization and some of the 
constraints required by such a change seem to differ again from those in 
ref. \cite{LuM92}. However, in this case, it was found in ref. \cite{FGM97} 
that to order $1/m_Q^2$ the difference is in a field redefinition so that 
to this order these two apparently disparate formulations are physically 
equivalent \cite{BaM97}. 

The above analysis is not limited to the effective Lagrangian --- 
it can equally well be applied to other reparameterization invariant 
combinations of operators such as currents. 
So next we turn to a similar analysis of the VRI predictions for the 
weak currents involving a heavy quark to order $1/m_Q^2$ which extends some 
previous results at order $1/m_Q$ \cite{LuM92,Neu94,Neu93}. We shall 
begin with an investigation of the heavy-light currents which in 
QCD are of the form $\bar q \, \Gamma^\alpha \, \psi_Q$ \cite{Neu94}. 
They are constructed from a heavy quark and a light quark field 
denoted generically by, $\psi_Q$ and $q$, respectively. These are 
the vector and axial-vector currents, respectively,
\alpheq
\eqa V^\alpha(x) = \bar q(x)\, \ga\, \psi_Q(x),
      \qquad (\Gamma^\alpha =\ga), \\
     \hat A^\alpha(x) = \bar q(x)\, \ga\, \gamma_5\, \psi_Q(x),
      \qquad (\Gamma^\alpha =\ga\, \gamma_5), 
\endeqa
\reseteq
which are clearly reparameterization invariant. As in the preceding analysis, 
the corresponding heavy quark fields in HQEFT will be represented by $Q_v$. 

Since the ensuing reparameterization analysis is virtually the same whether 
it is the vector or axial-vector current, we shall choose to treat the vector 
current in detail --- the corresponding results for the axial current 
are simply obtained by replacing $\bar q$ with $-\bar q \gamma_5$ 
\cite{FNL92}. While the vector current in QCD is conserved, radiative 
corrections in HQEFT not only effect current renormalization but can 
also induce additional operators not present at tree level. Hence the 
matrix element of the vector current in the full theory must be matched to 
a complete basis of current operators in the effective field theory with the 
same symmetries:
\eqa \bar q \, \ga\, \psi_Q &\simeq &
  \sum_i c_i^{(0)} (\mu) \, j_i^{(0)\alpha}
 +{1\over 2m_Q} \left[ \sum_j c_{1,j}^{(1)} (\mu) \, j_{1,j}^{(1)\alpha}
   +\sum_k c_{2,k}^{(1)} (\mu) \, j_{2,k}^{(1)\alpha} \right] \nonumber \\
 &&+{1\over (2m_Q)^2} \left[ \sum_l c_{1,l}^{(2)} (\mu) \, j_{1,l}^{(2)\alpha}
   +\sum_n c_{2,n}^{(2)} (\mu) \, j_{2,n}^{(2)\alpha} \right]
 +\ord{{1\over m_Q^3}},
\endeqal{HLVcurrexp}
where ``$\simeq$'' indicates an equality that holds only at the matrix 
element level, and $\mu$ is the renormalization scale.
In \eqr{HLVcurrexp}, a convenient operator basis is
\alpheq
%dim-3 operators
\eqa
 j_1^{(0)\alpha} &=& \bar{q}\, \ga\, Q_v, \nonumber \\
 j_2^{(0)\alpha} &=& \bar{q}\, v^\alpha\, Q_v, 
\endeqal{HLVdim3}
%dim-4 operators
\eq
 \begin{array}{ll}
  j_{1,1}^{(1)\alpha} = \bar{q}\, \ga\, i\Dslash\, Q_v, &\qquad
  j_{1,4}^{(1)\alpha} = -\bar{q}\, i\lvec{\Dslash}\, \ga\, Q_v, \\
  j_{1,1b}^{(1)\alpha} = \bar{q}\, \ga\, i\vD\, Q_v, &\qquad
  j_{1,5}^{(1)\alpha} = -\bar{q}\, i\lvec{\Dslash}\, v^\alpha\, Q_v, \\
  j_{1,2}^{(1)\alpha} = \bar{q}\, v^\alpha\, i\Dslash\, Q_v, &\qquad
  j_{1,6}^{(1)\alpha} = -\bar{q}\, iv\cdot\lvec{D}\, \ga\, Q_v, \\
  j_{1,2b}^{(1)\alpha} = \bar{q}\, v^\alpha\, i\vD\, Q_v, &\qquad
  j_{1,7}^{(1)\alpha} = -\bar{q}\, iv\cdot\lvec{D}\, v^\alpha\, Q_v, \\
  j_{1,3}^{(1)\alpha} = \bar{q}\, iD^\alpha\, Q_v, &\qquad
  j_{1,8}^{(1)\alpha} = -\bar{q}\, i\lvec{D^\alpha}\, Q_v,
 \end{array}
\endeql{HLVdim4}
% dim 4 T-ord'd prod
\eqa
 {j_{2,k}^{(1)\alpha}(x) \over 2m_Q} &=&
  i\int T\left[j_l^{(0)\alpha}(x) \,\oper{n}(y)\right]\, d^4y, \nonumber \\
  &&\qquad {\rm for}\; l=1,2,\> n=1a,1c,\> {\rm and}\, k=1,\ldots,4, 
\endeqal{HLVdim4Tprod}
%dim-5 local operators
\eq
 \begin{array}{ll}
  j_{1,1}^{(2)\alpha} = \bar{q}\, \ga\, (iD)^2\, Q_v, &\qquad
  j_{1,10}^{(2)\alpha} = -\bar{q}\, i\Dslashlft\, \ga\, i\Dslash\, Q_v, \\
  j_{1,2}^{(2)\alpha} = \bar{q}\, v^\alpha\, (iD)^2\, Q_v, &\qquad
  j_{1,11}^{(2)\alpha} = -\bar{q}\, i\Dslashlft\, v^\alpha\, i\Dslash\, Q_v, \\
  j_{1,3}^{(2)\alpha} = \bar{q}\, \ga\, (i\Dslash)^2\, Q_v, &\qquad
  j_{1,12}^{(2)\alpha} = -\bar{q}\, i\vDlft\, \ga\, i\Dslash\, Q_v, \\
  j_{1,4}^{(2)\alpha} = \bar{q}\, v^\alpha\, (i\Dslash)^2\, Q_v, &\qquad
  j_{1,13}^{(2)\alpha} = -\bar{q}\, i\vDlft\, v^\alpha\, i\Dslash\, Q_v, \\
  j_{1,5}^{(2)\alpha} = \bar{q}\, \ga\, i\vD\, i\Dslash\, Q_v, &\qquad
  j_{1,14}^{(2)\alpha} = -\bar{q}\, i\Dslashlft\, iD^\alpha\, Q_v, \\
  j_{1,5b}^{(2)\alpha} = \bar{q}\, \ga\, i\Dslash\, i\vD\, Q_v, &\qquad
  j_{1,15}^{(2)\alpha} = -\bar{q}\, i\lvec{D^\alpha}\, i\Dslash\, Q_v, \\
  j_{1,6}^{(2)\alpha} = \bar{q}\, v^\alpha\, i\vD\, i\Dslash\, Q_v, &\qquad
  j_{1,16}^{(2)\alpha} = -\bar{q}\, i\vDlft\, iD^\alpha\, Q_v, \\
  j_{1,6b}^{(2)\alpha} = \bar{q}\, v^\alpha\, i\Dslash\, i\vD\, Q_v, \\
  j_{1,7}^{(2)\alpha} = \bar{q}\, iD^\alpha\, i\Dslash\, Q_v, \\
  j_{1,7b}^{(2)\alpha} = \bar{q}\, iD^\alpha\, i\vD\, Q_v, \\
  j_{1,8}^{(2)\alpha} = \bar{q}\, i\Dslash\, iD^\alpha\, Q_v, \\
  j_{1,9}^{(2)\alpha} = \bar{q}\, i\vD\, iD^\alpha\, Q_v, \\
  j_{1,9b}^{(2)\alpha} = \bar{q}\, iD^\alpha\, i\vD\, Q_v, \\ \ldots, \\
 \end{array}
\endeql{HLVdim5}
% dim 5 T-ordered prod
\eq
 {j_{2,k}^{(2)\alpha}(x) \over 4m_Q^2} = \left\{
  \begin{array}{l}
  i\int T\left[j_l^{(0)\alpha}(x) \, \oper{n}(y)\right]\, d^4y, \\
   \qquad {\rm for}\; l=1,2,\> n=2a,2b,\> {\rm and}\, k=1,\ldots,4, \\
  \left({\delta_{mn}\over 2} -1 \right)
   \int T\left[j_l^{(0)\alpha}(x) \,\oper{m}(y),\oper{n}(z)\right]\, d^4y
    \,d^4z, \\
   \qquad {\rm for}\; l=1,2,\; (m,n)=(1a,1a),(1a,1c),(1c,1c),\; {\rm and}\,
    k=5,\ldots,10, \\
  i\int T\left[j_{1,l}^{(1)\alpha}(x) \,\oper{n}(y)\right]\, d^4y, \\
   \qquad {\rm for}\; l=1,\ldots,8,\; n=1a,1c,\; {\rm and}\, k=11,\ldots,26,
  \end{array}
  \right.
\endeql{HLVdim5Tprod}
\reseteq
where the ellipsis in \eqr{HLVdim5} denotes dimension-5 operators which are 
not related to the lower-dimensional ones through reparameterization 
invariance constraints. In \eqr{HLVdim5Tprod}, operators which vanish by the 
equation of motion, $i\vD\, Q_v =0$, 
have not been included because some of the time-ordered products involving 
such operators can be expressed in terms of other operators by contraction 
of internal heavy quark fields \cite{FLS94}; they have also been excluded 
from \eqr{HLVdim4Tprod}. Moreover, as explained next, reparameterization 
invariance does not further constrain time-ordered products of operators. 

In \eqr{HLVcurrexp}, the short-distance coefficients of those 
operators which are time-ordered products are easily obtained; they are 
just the product of the individual component operators which form it. So, 
for example, the coefficient of the dimension-4 operator 
$j_{2,1}^{(1)\alpha}$ with $k=1,\> l=1,\> n=1a,\>$ in \eqr{HLVdim4Tprod} 
above is 
\alpheq
\eq c_{2,1}^{(1)} = c_1^{(0)} \> b_{1a}, \nonumber \endeq
while for the dimension-5 operator $j_{2,4}^{(2)\alpha}$ with 
$k=4,\> l=2,\> n=2b$ in \eqr{HLVdim5Tprod} it is 
\eq c_{2,4}^{(2)} = c_2^{(0)} \> b_{2b}. \endeq
\reseteq

As for the effective Lagrangian above, the variation of the operators in this 
current expansion under a reparameterization transformation is required to 
vanish. When the variation is evaluated at the space-time origin for 
simplicity, this condition then yields the following constraints. 
\eqa
 c_1^{(0)} &=& c_{1,1}^{(1)} \nonumber\\
 2c_2^{(0)} = 2c_{1,2}^{(1)} &=& c_{1,3}^{(1)} \nonumber\\
 c_1^{(0)} -c_{1,1b}^{(1)} +c_{1,3}^{(2)} +c_{1,3b}^{(2)} &=& 0 \nonumber\\
 c_2^{(0)} -c_{1,2b}^{(1)} +c_{1,5}^{(2)} +c_{1,5b}^{(2)} &=& 0 \nonumber\\
 c_{1,1}^{(1)} +c_{1,1b}^{(1)} -c_{1,1}^{(2)} -c_{1,3}^{(2)} &=& 0 \nonumber\\
 2c_{1,2}^{(1)} -c_{1,7}^{(2)} -c_{1,8}^{(2)} &=& 0 \nonumber\\
 c_{1,2}^{(1)} +c_{1,2b}^{(1)} -c_{1,2}^{(2)} -c_{1,4}^{(2)} &=& 0 
  \label{HLVcurrRI}\\
 2c_{1,2b}^{(1)} -c_{1,7b}^{(2)} -c_{1,9}^{(2)} &=& 0 \nonumber\\
 c_{1,3}^{(1)} -c_{1,7}^{(2)} -c_{1,8}^{(2)} &=& 0 \nonumber\\
 c_{1,4}^{(1)} &=& c_{1,10}^{(2)} \nonumber\\
 2c_{1,5}^{(1)} = 2c_{1,11}^{(2)} &=& c_{1,14}^{(2)} \nonumber\\
 c_{1,6}^{(1)} &=& c_{1,12}^{(2)} \nonumber\\
 2c_{1,7}^{(1)} = 2c_{1,13}^{(2)} &=& c_{1,16}^{(2)} \nonumber\\
 c_{1,8}^{(1)} &=& c_{1,15}^{(2)} \nonumber
\endeqa
These relations are consistent with those obtained for the coefficients of 
the dimension-three and four operators in ref. \cite{Neu94}.

Since the equations in (\ref{HLVcurrRI}) relate the coefficients of 
lower-dimensional operators to those of higher dimension, one can then 
readily obtain the coefficient of the latter by performing the much simpler 
calculation of the former. For instance, the coefficients of the 
dimension-4 current operators $j_{1,i}^{(1)\alpha} \hbox{ for } i=6,7,8$,  
in \eqr{HLVdim4} have been constructed to next-to-leading order in 
renormalization-group-improved perturbation theory \cite{Neu94};
then the last three equations in (\ref{HLVcurrRI}) immediately determine the 
corresponding dimension-5 operators 
$j_{1,j}^{(2)\alpha} \hbox{ for } i=12,13,15,16$. Furthermore, the 
coefficient of the dimension-3 operator $j_1^{(0)\alpha}$ in \eqr{HLVdim3}, 
has been calculated to two loops in ref. \cite{BrG95}. Then the 
first relation in \eqr{HLVcurrRI} immediately furnishes the coefficient 
$c_{1,1}^{(1)}$ of the dimension-4 operator $j_{1,1}^{(1)\alpha}$ to the 
same level of accuracy. Alternatively, coefficients calculated by any 
other method are required to satisfy the VRI constraints, \eqr{HLVcurrRI}. 

Another class of weak currents in QCD we shall now consider are the 
heavy-heavy currents which in QCD are of the form 
$\bar \psi_{Q'}\, \Gamma^\alpha\, \psi_Q$ 
(with $\Gamma^\alpha =\ga \hbox{ or } \ga\, \gamma_5$) and may be 
flavour-conserving ($Q' = Q$) or flavour-changing ($Q' \ne Q$) \cite{Neu93}. 
Such currents are also manifestly reparameterization invariant in QCD. 
The heavy quark fields in the effective theory corresponding to $\psi_Q$ 
and $\psi_{Q'}$ will be represented by $Q_v$ and $Q'_{v'}$, respectively. 
The presence of the different velocities $v$ and $v'$ introduces a new 
variable $w = \vv'$ on which the coefficients of operators in the effective 
theory can depend on. As for the heavy-light case above, we shall focus on 
the vector current from which it is easy to obtain the corresponding axial 
vector results as discussed previously. 

Equating matrix elements of the vector current in QCD to a full operator 
set in HQEFT yields the expansion
\eqa \bar \psi_{Q'}\, \ga\, \psi_Q &\simeq &
  \sum_i d_i^{(0)}(\mu,w)\, J_i^{(0)\alpha}
 +\sum_j \left[ {d_{1,j}^{(1)}(\mu,w) \over 2m_Q}
  + {d_{1,j}^{(1)'}(\mu,w)\over 2m_{Q'}} \right] J_{1,j}^{(1)\alpha}
    \nonumber\\
 &&+\sum_k {d_{2,k}^{(1)}(\mu,w) \, J_{2,k}^{(1)\alpha} \over 2M} \nonumber\\
 &&+\sum_l \left[ {d_{1,l}^{(2)}(\mu,w) \over (2m_Q)^2}
    +{d_{1,l}^{(2)'}(\mu,w) \over 4m_Q m_{Q'}}
    +{d_{1,l}^{(2)''}(\mu,w) \over (2m_{Q'})^2} \right] J_{1,l}^{(2)\alpha}
    \nonumber\\
 &&+\sum_n \left[ d_{2,n}^{(2)}(\mu,w) \, J_{2,n}^{(2)\alpha}\over 4M^2 \right]
 +\ord{{1\over M^3}},
\endeqal{HHVcurrexp}
where $M = m_Q\> {\rm or}\> m_{Q'}$. 
A suitable basis for the operators in \eqr{HHVcurrexp} is
\alpheq
%dim-3 operators
\eq
 J_{1,2,3}^{(0)\alpha} = \bar{Q'}_{v'}\, \Omga \, Q_v, 
\endeql{HHVdim3}
%dim-4 operators
\eq
 \begin{array}{ll}
  J_{1,\{1,2,3\}}^{(1)\alpha} = \bar{Q'}_{v'}\, \Omga\, i\Dslash\, Q_v, &\qquad
  J_{1,\{8,9,10\}}^{(1)\alpha} = -\bar{Q'}_{v'}\, i\Dslashlft\, \Omga\, Q_v, \\
  J_{1,\{4,5,6\}}^{(1)\alpha} = \bar{Q'}_{v'}\, \Omga\, i\vpD\, Q_v, &\qquad
  J_{1,\{11,12,13\}}^{(1)\alpha} = -\bar{Q'}_{v'}\, i\vDlft\, \Omga\, Q_v, \\
  J_{1,\{4b,5b,6b\}}^{(1)\alpha} = \bar{Q'}_{v'}\, \Omga\, i\vD\, Q_v, &\qquad
  J_{1,\{11b,12b,13b\}}^{(1)\alpha} =-\bar{Q'}_{v'}\, i\v'Dlft\, \Omga\, Q_v,\\
  J_{1,7}^{(1)\alpha} = \bar{Q'}_{v'}\, iD^\alpha\, Q_v, &\qquad
  J_{1,14}^{(1)\alpha} = -\bar{Q'}_{v'}\, i\lvec{D^\alpha}\, Q_v, 
 \end{array}
\endeql{HHVdim4}
% dim-4 T-ord'd prod
\eq
 {J_{2,k}^{(1)\alpha}(x) \over 2M} = \left\{
  \begin{array}{l}
  i\int T\left[J_l^{(0)\alpha}(x) \,\oper{n}(y)\right] d^4y, \nonumber \\
   \qquad {\rm for}\; M=m_Q,\> l=1,2,3,\> n=1a,1c,\> {\rm and}\,k=1,\ldots,6,\\
  i\int T\left[J_l^{(0)\alpha}(x) \,\oper{n}'(y)\right] d^4y, \nonumber \\
   \qquad {\rm for}\; M=m_{Q'},\> l=1,2,3,\> n=1a,1c,\> {\rm and}\,
   k=7,\ldots,12, \\
  \end{array}
  \right.
\endeql{HHVdim4Tprod}
% dim-5 operators
\eq
 \begin{array}{ll}
  J_{1,\{1,2,3\}}^{(2)\a} = \bQ'v'\, \Omega^\a\, (iD)^2\, Q_v, &\qquad
  J_{1,\{24,25,26\}}^{(2)\a} = \bQ'v'\,(i\lvec{D})^2\,\Omega^\a\, Q_v,\\
  J_{1,\{4,5,6\}}^{(2)\a} = \bQ'v'\, \Omega^\a\, (i\Dslash)^2\, Q_v, &\qquad
  J_{1,\{27,28,29\}}^{(2)\a} = \bQ'v'\, (i\lvec{\Dslash})^2\,
   \Omega^\a\, Q_v, \\
  J_{1,\{7,8,9\}}^{(2)\a} = \bQ'v'\, \Omega^\a\, i\vD\,i\Dslash\, Q_v, &\qquad
  J_{1,\{30,31,32\}}^{(2)\a} = \bQ'v'\, i\Dslashlft\, i\vDlft\,
    \Omega^\a\, Q_v, \\
  J_{1,\{10,11,12\}}^{(2)\a} = \bQ'v'\,\Omega^\a\,i\vpD\,i\Dslash\, Q_v,&\qquad
  J_{1,\{33,34,35\}}^{(2)\a} = \bQ'v'\, i\Dslashlft\, i\v'Dlft\,
    \Omega^\a\, Q_v, \\
  J_{1,\{13,14,15\}}^{(2)\a} = \bQ'v'\,\Omega^\a\,i\Dslash\,i\vpD\, Q_v,&\qquad
  J_{1,\{36,37,38\}}^{(2)\a} = \bQ'v'\, i\vDlft\, i\Dslashlft\,
    \Omega^\a\, Q_v, \\
  J_{1,\{13b,14b,15b\}}^{(2)\a} =\bQ'v'\,\Omega^\a\,i\Dslash\,i\vD\,Q_v,&\qquad
  J_{1,\{36b,37b,38b\}}^{(2)\a} = \bQ'v'\, i\v'Dlft\, i\Dslashlft\,
    \Omega^\a\, Q_v, \\
  J_{1,\{16,17,18\}}^{(2)\a} = \bQ'v'\,\Omega^\a\, (i\vpD)^2\, Q_v, &\qquad
  J_{1,\{39,40,41\}}^{(2)\a} = \bQ'v'\,(i\vDlft)^2\, \Omega^\a\, Q_v,\\
  J_{1,\{16b,17b,18b\}}^{(2)\a} =\bQ'v'\,\Omega^\a\, i\vpD\, i\vD\, Q_v,&\qquad
  J_{1,\{39b,40b,41b\}}^{(2)\a} =\bQ'v'\, i\v'Dlft\,i\vDlft\,\Omega^\a\, Q_v,\\
  J_{1,\{16c,17c,18c\}}^{(2)\a} =\bQ'v'\,\Omega^\a\, i\vD\, i\vpD\, Q_v,&\qquad
  J_{1,\{39c,40c,41c\}}^{(2)\a} =\bQ'v'\, i\vDlft\,i\v'Dlft\,\Omega^\a\, Q_v,\\
  J_{1,19}^{(2)\a} = \bQ'v'\, iD^\a\, i\Dslash\, Q_v, &\qquad
  J_{1,42}^{(2)\a} = \bQ'v'\, i\lvec{\Dslash}\, i\lvec{D^\a}\, Q_v, \\
  J_{1,20}^{(2)\a} = \bQ'v'\, iD^\a\, i\vpD\, Q_v, &\qquad
  J_{1,43}^{(2)\a} = \bQ'v'\, i\vDlft\, i\lvec{D^\a}\, Q_v, \\
  J_{1,20b}^{(2)\a} = \bQ'v'\, iD^\a\, i\vD\, Q_v, &\qquad
  J_{1,43b}^{(2)\a} = \bQ'v'\, i\v'Dlft\, i\lvec{D^\a}\, Q_v, \\
  J_{1,21}^{(2)\a} = \bQ'v'\, i\Dslash\, iD^\a\, Q_v, &\qquad
  J_{1,44}^{(2)\a} = \bQ'v'\, i\lvec{D^\a}\, i\lvec{\Dslash}\, Q_v, \\
  J_{1,22}^{(2)\a} = \bQ'v'\, i\vD\, iD^\a\, Q_v, &\qquad
  J_{1,45}^{(2)\a} = \bQ'v'\, i\lvec{D^\a}\, i\vDlft\, Q_v, \\
  J_{1,23}^{(2)\a} = \bQ'v'\, i\vpD\, iD^\a\, Q_v, &\qquad
  J_{1,46}^{(2)\a} = \bQ'v'\, i\lvec{D^\a}\, i\v'Dlft\, Q_v, \\
  J_{1,\{47,48,49\}}^{(2)\a} = -\bQ'v'\, i\lvec{\Dslash}\, \Omega^\a\,
   i\Dslash\, Q_v, \\
  J_{1,\{50,51,52\}}^{(2)\a} = -\bQ'v'\, i\lvec{\Dslash}\, \Omega^\a\,
   i\vpD\, Q_v, &\qquad
  J_{1,\{53,54,55\}}^{(2)\a} = -\bQ'v'\, i\vDlft\, \Omega^\a\, i\Dslash\,Q_v,\\
  J_{1,\{50b,51b,52b\}}^{(2)\a} = -\bQ'v'\, i\lvec{\Dslash}\, \Omega^\a\,
   i\vD\, Q_v, &\qquad
  J_{1,\{53b,54b,55b\}}^{(2)\a} = -\bQ'v'\, i\v'Dlft\, \Omega^\a\, i\Dslash\,
   Q_v,\\
  J_{1,56}^{(2)\a} = -\bQ'v'\, i\lvec{\Dslash}\, iD^\a\, Q_v, &\qquad
  J_{1,57}^{(2)\a} = -\bQ'v'\, i\lvec{D^\a}\, i\Dslash\, Q_v, \\
  J_{1,\{58,59,60\}}^{(2)\a} = -\bQ'v'\, \Omga i\lvec{D}\cdot iD\, Q_v, \\
  J_{1,\{61,62,63\}}^{(2)\a} = -\bQ'v'\,\Omega^\a\, i\vDlft\, i\vpD\,
   Q_v, \\
  J_{1,64}^{(2)\a} = -\bQ'v'\, i\vDlft\, iD^\a Q_v, &\qquad
  J_{1,65}^{(2)\a} = -\bQ'v'\, i\lvec{D^\a}\, i\vpD\, Q_v, \\
  J_{1,64b}^{(2)\a} = -\bQ'v'\, i\v'Dlft\, iD^\a Q_v, &\qquad
  J_{1,65b}^{(2)\a} = -\bQ'v'\, i\lvec{D^\a}\, i\vD\, Q_v, \\
  \ldots,
 \end{array}
\endeql{HHVdim5}
% dim-5 T-ordered prod
\eq
 {J_{2,k}^{(2)\alpha}(x) \over 4M^2} = \left\{
  \begin{array}{l}
  i\int T\left[J_l^{(0)\alpha}(x) \,\oper{n}(y)\right]\, d^4y, \\
   \qquad {\rm for}\; M^2=m_Q^2,\> l=1,2,3,\> n=2a,2b,\> {\rm and}\,
    k=1,\ldots,6,\\
  i\int T\left[J_l^{(0)\alpha}(x) \,\oper{n}'(y)\right]\, d^4y, \\
   \qquad {\rm for}\; M^2=m_{Q'}^2,\> l=1,2,3,\> n=2a,2b,\> {\rm and}\,
    k=7,\ldots,12,\\
  \left({\delta_{mn}\over 2} -1 \right)
   \int T\left[J_l^{(0)\alpha}(x)\,\oper{m}(y)\,\oper{n}(z)\right]\, d^4y\,
    d^4z,\\
   \qquad {\rm for}\; M^2 =m_Q^2,\> l=1,2,3,\; (m,n)=(1a,1a),(1a,1c),(1c,1c),\\
    \qquad {\rm and}\, k=13,\ldots,21, \nonumber \\
  -\int T\left[J_l^{(0)\alpha}(x) \,\oper{m}(y) \,\oper{n}'(z)\right]\, d^4y\,
    d^4z, \\
   \qquad {\rm for}\; M^2 =m_Q m_{Q'},\> l=1,2,3,\; (m,n)=(1a,1a),(1a,1c),
    (1c,1a),\\
   \qquad (1c,1c),{\rm and}\, k=22,\ldots,33, \\
  \left({\delta_{mn}\over 2} -1 \right)
   \int T\left[J_l^{(0)\alpha}(x)\, \oper{m}'(y)\, \oper{n}'(z)\right]\, d^4y 
    \,d^4z, \\
   \qquad {\rm for}\; M^2 =m_{Q'}^2,\> l=1,2,3,\; (m,n)=(1a,1a),(1a,1c),
    (1c,1c),\\ \qquad {\rm and}\, k=34,\ldots,42, \nonumber \\
  i\int T\left[J_{1,l}^{(1)\alpha}(x)\, \oper{n}(y)\right]\, d^4y, \\
   \qquad {\rm for}\; M^2 = m_Q^2\> {\rm or}\> m_Q m_{Q'},\; l=1,\ldots,14,\;
    n=1a,1c, \\\qquad {\rm and}\, k=43,\ldots,70, \\
  i\int T\left[J_{1,l}^{(1)\alpha}(x)\,\oper{n}'(y)\right]\, d^4y, \\
   \qquad {\rm for}\; M^2 = m_{Q'}^2\> {\rm or}\> m_Q m_{Q'},\;
    l=1,\ldots,14,\; n=1a,1c, \\\qquad {\rm and}\, k=71,\ldots,98, \\
  \end{array}
  \right.
\endeql{HHVdim5Tprod}
\reseteq
where $\Omega^\a = \{\ga,\va,\vpa\}$ so that the operators in 
eq. (\ref{HHVdim3}), (\ref{HHVdim4}) and (\ref{HHVdim5}) denoted by the 
symbol $J$ with a second subscript consisting of 
a triplet of numbers in curly braces correspond to the triplet of entries 
in $\Omega^\a$, respectively. Hence, for example, $J_{1,1}^{(2)\a}$ is the 
operator containing $\ga$, $J_{1,2}^{(2)\a}$ contains $\va$, and 
$J_{1,3}^{(2)\a}$ contains $\vpa$. The operators $\{\oper{n}\}$ were given 
in \eqr{LHQEFTOp} while the set $\{\oper{n}'\}$ is defined analogously by 
the same equations except with the replacements 
$Q_v \rightarrow Q'_{v'}$ and $m_Q \rightarrow m_{Q'}$. 
The ellipsis in \eqr{HHVdim5} represents dimension-5 operators which are not 
constrained by lower-dimensional ones through reparameterization invariance. 
As in the heavy-light analysis, time-ordered products involving operators 
which vanish by the equation of motion have been omitted from 
\eqr{HHVdim4Tprod} and (\ref{HHVdim5Tprod}). 

However, in contrast to the heavy-light case there is a subtlety which did 
not arise before. Since the short-distance coefficients here also depend on 
the velocities through the quantity $w =\vv'$, they will also transform under 
a change in the velocity: under the velocity shift $v \rightarrow v +\Delta v$,
\alpheq
\eq d(\vv') \rightarrow d(\vv') +(\Delta v \cdot v') {\del d(w)\over \del w}
  +\ord{(\Delta v)^2},
\endeq
while for $v' \rightarrow v' +\Delta v'$,
\eq d(\vv') \rightarrow d(\vv') +(v \cdot \Delta v') {\del d(w)\over \del w}
  +\ord{(\Delta v')^2},
\endeq
\reseteq
The other arguments as well as the subscripts and superscripts of the 
coefficients have been suppressed here. Hence the right-hand side of the 
heavy-heavy current expansion, \eqr{HHVcurrexp}, is reparameterization 
invariant when not only the operators but also the arguments of the 
coefficients are transformed under a shift in the velocities and the 
residual momenta. This approach is equivalent to the observation in ref. 
\cite{Neu93} that both the operators and the coefficients must be written 
in a form which is invariant under reparameterization. Applying this 
requirement to \eqr{HHVcurrexp} and working to linear order in the shifts 
in the velocity and residual momentum and to order $1/M$ yields the 
following constraints. 

\alpheq
\eqa
 &
 \begin{array}{ll}
  d_{1,j}^{(1)} = 0, &\qquad {\rm for}\> j = 8,\ldots,14, \\
  d_{1,j}^{(1)'} = 0, &\qquad {\rm for}\> j = 1,\ldots,7, \\
  d_{1,j}^{(2)} = 0, &\qquad {\rm for}\> j = 24,\ldots,65b, \\
  d_{1,j}^{(2)'} = 0, &\qquad {\rm for}\> j = 1,\ldots,46, \\
  d_{1,j}^{(2)''} = 0, &\qquad {\rm for}\> j = 1,\ldots,23,\> {\rm and}\>
   47,\ldots,65b. \\
 \end{array}\label{RIderiv}
 &
 \\\noalign{\vskip 3mm}
 &\hskip -1.5em
 \begin{array}{l}
  d_1^{(0)} = d_{1,1}^{(1)} = d_{1,8}^{(1)'} = d_{1,47}^{(2)'} \\
  2d_2^{(0)} = 2d_{1,2}^{(1)} = d_{1,7}^{(1)} = 2d_{1,9}^{(1)'}
   = 2d_{1,48}^{(2)'} = d_{1,56}^{(2)'} \\
  2d_3^{(0)} = 2d_{1,3}^{(1)} = 2d_{1,10}^{(1)'} = d_{1,14}^{(1)'}
   = 2d_{1,49}^{(2)'} = d_{1,57}^{(2)'} \\
  2d_{1,4}^{(1)} = 2d_{1,11}^{(1)'} = 2d_{1,50}^{(2)'} = 2d_{1,53}^{(2)'}
   = d_{1,58}^{(2)'} \\
  2d_{1,5}^{(1)} = 2d_{1,12}^{(1)'} = 2d_{1,51}^{(2)'} = 2d_{1,54}^{(2)'}
   = d_{1,59}^{(2)'} = d_{1,64}^{(2)'} \\
  2d_{1,6}^{(1)} = 2d_{1,13}^{(1)'} = 2d_{1,52}^{(2)'} = 2d_{1,55}^{(2)'}
   = d_{1,60}^{(2)'} = d_{1,65}^{(2)'} \\
  2{\del d_1^{(0)}\over \del w} = d_{1,4}^{(1)} = d_{1,11}^{(1)'}
   = d_{1,53}^{(2)'} \\
  2{\del d_2^{(0)}\over \del w} = d_{1,5}^{(1)} = d_{1,12}^{(1)'}
   = d_{1,54}^{(2)'} \\
  2{\del d_3^{(0)}\over \del w} = d_{1,6}^{(1)} = d_{1,13}^{(1)'}
   = d_{1,55}^{(2)'} \\
 \end{array}\label{RIm^0,1/mor1/m'}
 &
 \\
 &
 \begin{array}{ll}
  d_1^{(0)} -d_{1,4b}^{(1)} +d_{1,7}^{(2)} +d_{1,13b}^{(2)} = 0 &\qqquad
  d_1^{(0)} -d_{1,11b}^{(1)'} +d_{1,30}^{(2)''} +d_{1,36b}^{(2)''} = 0 \\
  d_{1,1}^{(1)} +d_{1,4b}^{(1)} -d_{1,1}^{(2)} -d_{1,4}^{(2)} = 0 &\qqquad
  d_{1,8}^{(1)'} +d_{1,11b}^{(1)'} -d_{1,24}^{(2)''} -d_{1,27}^{(2)''} = 0 \\
  d_2^{(0)} -d_{1,5b}^{(1)} +d_{1,8}^{(2)} +d_{1,14b}^{(2)} = 0 &\qqquad
  d_2^{(0)} -d_{1,12b}^{(1)'} +d_{1,31}^{(2)''} +d_{1,37}^{(2)''} = 0 \\
  2d_{1,2}^{(1)} -d_{1,19}^{(2)} -d_{1,21}^{(2)} = 0 &\qqquad
  2d_{1,9}^{(1)'} -d_{1,42}^{(2)''} -d_{1,44}^{(2)''} = 0 \\
  d_{1,2}^{(1)} +d_{1,5b}^{(1)} -d_{1,2}^{(2)} -d_{1,5}^{(2)} = 0 &\qqquad
  d_{1,9}^{(1)'} +d_{1,12b}^{(1)'} -d_{1,25}^{(2)''} -d_{1,28}^{(2)''} = 0 \\
  2d_{1,5b}^{(1)} -d_{1,20b}^{(2)} -d_{1,22}^{(2)} = 0 &\qqquad
  2d_{1,12b}^{(1)'} -d_{1,43b}^{(2)''} -d_{1,45}^{(2)''} = 0 \\
  d_3^{(0)} -d_{1,6b}^{(1)} +d_{1,9}^{(2)} +d_{1,15b}^{(2)} = 0 &\qqquad
  d_3^{(0)} -d_{1,13b}^{(1)'} +d_{1,32}^{(2)''} +d_{1,38b}^{(2)''} = 0 \\
  d_{1,3}^{(1)} +d_{1,6b}^{(1)} -d_{1,3}^{(2)} -d_{1,6}^{(2)} = 0 &\qqquad
  d_{1,10}^{(1)'} +d_{1,13b}^{(1)'} -d_{1,26}^{(2)''} -d_{1,29}^{(2)''} = 0 \\
  d_{1,4}^{(1)} -d_{1,10}^{(2)} -d_{1,13}^{(2)} = 0 &\qqquad
  d_{1,11}^{(1)'} -d_{1,33}^{(2)''} -d_{1,36}^{(2)''} = 0 \\
  2d_{1,5}^{(1)} -d_{1,20}^{(2)} -d_{1,23}^{(2)} = 0 &\qqquad
  2d_{1,12}^{(1)'} -d_{1,43}^{(2)''} -d_{1,46}^{(2)''} = 0 \\
  d_{1,5}^{(1)} -d_{1,11}^{(2)} -d_{1,14}^{(2)} = 0 &\qqquad
  d_{1,12}^{(1)'} -d_{1,34}^{(2)''} -d_{1,37}^{(2)''} = 0 \\
  d_{1,6}^{(1)} -d_{1,12}^{(2)} -d_{1,15}^{(2)} = 0 &\qqquad
  d_{1,13}^{(1)'} -d_{1,35}^{(2)''} -d_{1,38}^{(2)''} = 0 \\
  d_{1,7}^{(1)} -d_{1,19}^{(2)} -d_{1,21}^{(2)} = 0 &\qqquad
  d_{1,14}^{(1)'} -d_{1,42}^{(2)''} -d_{1,44}^{(2)''} = 0 \\
  2{\del d_{1,1}^{(1)}\over\del w} -d_{1,10}^{(2)} -d_{1,13}^{(2)} = 0 &\qqquad
  2{\del d_{1,8}^{(1)'}\over\del w} -d_{1,33}^{(2)''} -d_{1,36}^{(2)''} = 0 \\
  2{\del d_{1,2}^{(1)}\over\del w} -d_{1,11}^{(2)} -d_{1,14}^{(2)} = 0 &\qqquad
  2{\del d_{1,9}^{(1)'}\over\del w} -d_{1,34}^{(2)''} -d_{1,37}^{(2)''} = 0 \\
  2{\del d_{1,3}^{(1)}\over\del w} -d_{1,12}^{(2)} -d_{1,15}^{(2)} = 0 &\qqquad
  2{\del d_{1,10}^{(1)'}\over\del w} -d_{1,35}^{(2)''} -d_{1,38}^{(2)''} = 0 \\
  {\del d_{1,4}^{(1)}\over\del w} -d_{1,16}^{(2)} = 0 &\qqquad
  {\del d_{1,11}^{(1)'}\over\del w} -d_{1,39}^{(2)''} = 0 \\
  2{\del d_{1,4b}^{(1)}\over\del w} -d_{1,16b}^{(2)} -d_{1,16c}^{(2)} = 0
   &\qqquad
  2{\del d_{1,11b}^{(1)'}\over\del w} -d_{1,39b}^{(2)''} -d_{1,39c}^{(2)''}=0\\
  {\del d_{1,5}^{(1)}\over\del w} -d_{1,17}^{(2)} = 0 &\qqquad
  {\del d_{1,12}^{(1)'}\over\del w} -d_{1,40}^{(2)''} = 0 \\
  2{\del d_{1,5b}^{(1)}\over\del w} -d_{1,17b}^{(2)} -d_{1,17c}^{(2)} = 0
   &\qqquad
  2{\del d_{1,12b}^{(1)'}\over\del w} -d_{1,40b}^{(2)''} -d_{1,40c}^{(2)''}=0\\
  {\del d_{1,6}^{(1)}\over\del w} -d_{1,18}^{(2)} = 0 &\qqquad
  {\del d_{1,13}^{(1)'}\over\del w} -d_{1,41}^{(2)''} = 0 \\
  2{\del d_{1,6b}^{(1)}\over\del w} -d_{1,18b}^{(2)} -d_{1,18c}^{(2)} = 0
   &\qqquad
  2{\del d_{1,13b}^{(1)'}\over\del w} -d_{1,41b}^{(2)''} -d_{1,41c}^{(2)''}=0\\
  2{\del d_{1,7}^{(1)}\over\del w} -d_{1,20}^{(2)} -d_{1,23}^{(2)} = 0 &\qqquad
  2{\del d_{1,14}^{(1)'}\over\del w} -d_{1,43}^{(2)''} -d_{1,46}^{(2)''} = 0 \\
 \end{array}\label{RI1/m}
 &
 \\\noalign{\vskip 3mm}
 &\hskip -1.3cm
 \begin{array}{ll}
  d_{1,4b}^{(1)} = d_{1,50b}^{(2)'} &\hskip 4.5cm 
   d_{1,11b}^{(1)'} = d_{1,53b}^{(2)'} \\
  d_{1,5b}^{(1)} = d_{1,51b}^{(2)'} &\hskip 4.5cm 
   2d_{1,12b}^{(1)'} = 2d_{1,54b}^{(2)'} = d_{1,64b}^{(2)'} \\
  d_{1,6b}^{(1)} = d_{1,52b}^{(2)'} &\hskip 4.5cm 
   d_{1,13b}^{(1)'} = d_{1,55b}^{(2)'} \\
  d_{1,7}^{(1)} = d_{1,56}^{(2)'} &\hskip 4.5cm 
   d_{1,14}^{(1)'} = d_{1,57}^{(2)'} \\
  2{\del d_{1,7}^{(1)}\over\del w} = d_{1,64}^{(2)'} &\hskip 4.5cm
   2{\del d_{1,8}^{(1)'}\over\del w} = d_{1,50}^{(2)'} \\
  2{\del d_{1,9}^{(1)'}\over\del w} = d_{1,51}^{(2)'} &\hskip 4.5cm
   2{\del d_{1,10}^{(1)'}\over\del w} = d_{1,52}^{(2)'} \\
  2{\del d_{1,14}^{(1)'}\over\del w} = d_{1,65}^{(2)'} \\
 \end{array}\label{HHVcurrRI4}
 &\\
 \nonumber\\
 &
 \begin{array}{r}
  2{\del d_{1,4}^{(1)}\over\del w} = 2{\del d_{1,11}^{(1)'}\over\del w} 
   = d_{1,61}^{(2)'} \\
  2{\del d_{1,4b}^{(1)}\over\del w} = 2{\del d_{1,11b}^{(1)'}\over\del w} 
   = d_{1,61b}^{(2)'} \\
  2{\del d_{1,5}^{(1)}\over\del w} = 2{\del d_{1,12}^{(1)'}\over\del w} 
   = d_{1,62}^{(2)'} \\
  2{\del d_{1,5b}^{(1)}\over\del w} = 2{\del d_{1,12b}^{(1)'}\over\del w} 
   = d_{1,62b}^{(2)'} \\
  2{\del d_{1,6}^{(1)}\over\del w} = 2{\del d_{1,13}^{(1)'}\over\del w} 
   = d_{1,63}^{(2)'} \\
  2{\del d_{1,6b}^{(1)}\over\del w} = 2{\del d_{1,13b}^{(1)'}\over\del w} 
   = d_{1,63b}^{(2)'} \\
 \end{array}\label{HHVcurrRI5}
 &
\endeqa
\reseteq

As for the heavy-light current, the coefficients of the operators given 
in \eqr{HHVdim4Tprod} and (\ref{HHVdim5Tprod}) whose definition involve 
time-ordered products are simply the product of the individual component 
operators forming them, and hence will not be displayed explicitly.

The VRI predictions given in ref. \cite{LuM92,Neu93} for the coefficients 
of some of the dimension-3 and dimension-4 operators in \eqr{HHVdim3} and 
(\ref{HHVdim4}), respectively, are generalized by the relations displayed 
in \eqr{RIderiv} and (\ref{RIm^0,1/mor1/m'}) above. 

Although the complete basis of current operators on the right-hand side of 
\eqr{HHVcurrexp} grows dramatically at higher orders in $1/M$, the relations 
required by VRI provide us with considerable predictive power. As pointed 
out in ref. \cite{Neu93}, reparameterization invariance relates the 
coefficients of all local dimension-4 operators which do not vanish by the 
equations of motion in the current expansion, \eqr{HHVcurrexp}, to those of 
the dimension-3 operators; the precise conditions are given by a subset of 
\eqr{RIderiv} and (\ref{RIm^0,1/mor1/m'}). While the coefficients of the 
dimension-5 operators in \eqr{HHVcurrexp} cannot all be uniquely determined 
in terms of lower-dimensional ones solely on the basis of VRI, it provides 
a large number of exact constraints which are displayed above in eq. 
(\ref{RIderiv}-\ref{HHVcurrRI5}). 

Some dimension-5 operators, however, are completely determined by these 
equations as we shall now exemplify. The coefficients of the dimension-3 
operators, $J_i^{(0)\alpha} \hbox{ for } i=1,2,3$, and the dimension-4 
operators, $J_i^{(1)\alpha} \hbox{ for } i=\hbox{1-6,8-13}$, were calculated 
in the leading logarithmic approximation in ref. \cite{FNL92}. The relations 
in \eqr{RIm^0,1/mor1/m'}, (\ref{HHVcurrRI4}), and (\ref{HHVcurrRI5}) will then 
uniquely determine the coefficients of the dimension-5 operators in these 
equations in terms of them. 
For instance, the second relation in \eqr{RIm^0,1/mor1/m'} gives 
\eq 2d_2^{(0)} = 2d_{1,2}^{(1)} = d_{1,7}^{(1)} = 2d_{1,9}^{(1)'}
   = 2d_{1,48}^{(2)'} = d_{1,56}^{(2)'}
   = \left[ {\alpha_s(\bar m) \over \alpha_s(\mu)} \right]^{a_L(w)}, \endeq
while using this result in combination with the eighth relation in 
\eqr{RIm^0,1/mor1/m'} yields
\eq 2{\del d_2^{(0)}\over \del w} = d_{1,5}^{(1)} = d_{1,12}^{(1)'} 
   = d_{1,54}^{(2)'}
   = 2 d_2^{(0)} {\del a_L(w)\over \del w}
      \ln\left[ {\alpha_s(\bar m) \over \alpha_s(\mu)} \right]^{a_L(w)}, \endeq
where 
\eqa a_L(w) = {8 \over 33 -2n_f} [w\, r(w) -1], \\
     r(w) = {\ln (w +\sqrt{w^2 -1}) \over \sqrt{w^2 -1}}. \nonumber \endeqa
The quantity $\bar m$ is an average mass between $m_Q$ and $m_{Q'}$.
Furthermore, from the first line of \eqr{HHVcurrRI5} one obtains
\eq 2{\del d_{1,4}^{(1)}\over\del w} = 2{\del d_{1,11}^{(1)'}\over\del w} 
   = d_{1,61}^{(2)'} 
   =-{8\over 33-2n_f} \left[\ln {\alpha_s(\bar m) \over \alpha_s(\mu)} \right]
     {\del\over \del w} \left\{{r(w) -1 \over w^2 -1}
     \left[\ln {\alpha_s(\bar m) \over \alpha_s(\mu)}\right]^{a_L(w)} \right\}.
\endeq
In these equations, the dependence of the coefficients on $\mu$ and $w$ have 
been omitted for simplicity. With relations such as these, once the 
coefficients of the dimension-3 operators have been calculated to subleading 
order, the coefficients of not only some dimension-4 but also some 
dimension-5 operators are immediately determined to the same accuracy without 
additional labour. 

In the next section, we shall similarly treat systems of higher spin which 
involve constraints. 

\section{Reparameterization Invariance of a Heavy Vector Effective Field
 Theory}

To illustrate the application of the above methods to a massive spin-1 
system, we shall use the model field theory previously considered in 
ref. \cite{Lee97a} with the Lagrangian
\eq \Lagr_V = -\frac{1}{2} 
               (D_\mu A_\nu -D_\nu A_\mu)^\dagger (D^\mu A^\nu -D^\nu A^\mu) 
              +(m_V)^2 A_\mu^\dagger A^\mu.\endeq
In $\Lagr_V$, $A^\mu$ is the vector field with mass $m_V$ with interactions 
prescribed by the gauge-covariant derivative $D^\mu$. As we saw in ref. 
\cite{Lee97a}, the effective field theory for this system can be expressed 
in terms of the dynamical field $\Avperpl{\mu}$ which acts only on vectors 
and not antivectors. It is related to the field in the full theory $A^\mu$ 
through
\eqa
 \Avperpl{\mu}
  &=& \left(1+{i\vD\over 2m_V}\right) e^{im_V v\cdot x} A_\mu^\perp,\\
 A_\mu^\perp &=& (g_{\mu\nu} -v_\mu v_\nu) A^\nu, \nonumber
\endeqa
and its transformation under a velocity reparameterization \cite{Lee97a} is
determined by
\alpheq
\eqa \Avpl{\mu} \rightarrow \tilde A_{\mu, v}^+ \hskip -.5em
 &=& \hskip -.5em e^{im_V\Delta v \cdot x}
     \left[\Avpl{\mu} +\frac{i\Delta v\cdot D}{2m_V} A_{\mu,v}
      +\ord{(\Delta v)^2}\right] \label{AvplVRexact}\\
 &=& \hskip -.5em e^{im_V\Delta v \cdot x}
     \left\{\Avpl{\mu} +{i\Delta v \cdot D \over 2m_V}
     \left[\Avpl{\mu} - (B_2^{-1})_\mu{}^\nu C_\nu \right] +\ord{(\Delta v)^2}
      \right\} \\
 \hskip -.5em
 &=& \hskip -.5em e^{im_V\Delta v \cdot x}
     \Biggl\{\Avpl{\mu} +{i\Delta v \cdot D \over 2m_V}
     \left[\Avpl{\mu} -\left(\frac{1}{4(m_V)^2}
      -\frac{i\vD}{8(m_V)^3}\right)C_\mu \right] \nonumber\\
 && \hskip 4em +\ord{(\Delta v)^2, \smlfrac{\Delta v}{(m_V)^5}} \Biggr\},
     \label{AvplVR}
\endeqa
\reseteq
where
\eqa (B_2)_{\mu\nu}
 &=& [2m_V(2m_V +i\vD) +(\Dperp)^2]g_{\mu\nu} -D_{\perp\nu} D_{\perp\mu}
  \nonumber\\
 && -(\vD -im_V)D_{\perp\mu} [(\Dperp)^2 +(m_V)^2]^{-1} D_{\perp\nu}
  (\vD -im_V),
\endeqa
and
\eq C_\nu
 = (\Dperp)^2 \Avpl{\nu} -D_\perp^\alpha D_{\perp\nu}\Avpl{\alpha}
  -(\vD -im_V)D_{\perp\nu}[(\Dperp)^2 +(m_V)^2]^{-1} D_\perp^\alpha (\vD -im_V)
          \Avpl{\alpha}. \endeq
The exact transformation is given by \eqr{AvplVRexact}, but in the two 
equalities following it, a result from the path integral calculation in 
ref. \cite{Lee97a} has once again been used to write $A_\mu^-$ in terms of 
$\Avpl{\mu}$. 

As in the other cases examined above, we shall determine the implications of 
these transformations on the most general set of operators which respect the 
symmetries of the theory and hence may appear in the Lagrangian:
\eq \Lagr_{\rm HVEFT}^{\rm gen} = \sum_v \L{V}{},\endeq
where
\eq \L{V}{} = \sum_j \frac{\L{V}{(j)}}{(m_V)^{j-1}}, \endeq
in which the first five terms in this expansion are
\alpheq
\eqa
 \L{V}{(0)} &=& -2g^{\mu\nu} (\Avperpl{\mu})^\dagger i\vD \Avperpl{\nu},\\
 \L{V}{(1)} &=& (\Avperpl{\mu})^\dagger
   \left\{g^{\mu\nu} \left[d_{1a} D^2 +d_{1b} (\vD)^2 \right]
   +d_{1c} D^\mu D^\nu +d_{1d} D^\nu D^\mu \right\} \Avperpl{\nu},\\
 \L{V}{(2)} &=& i(\Avperpl{\mu})^\dagger 
    \left\{ g^{\mu\nu} \left[d_{2a} D_\alpha \vD D^\alpha 
     +d_{2b} (D^2 \vD +\vD D^2) +d_{2c} (\vD)^3 \right] \right.\nonumber\\
  &&+d_{2d} (D^\mu D^\nu \vD + \vD D^\mu D^\nu) 
    +d_{2e} (D^\nu D^\mu \vD +\vD D^\nu D^\mu ) \nonumber \\
  &&+d_{2f} D^\mu \vD D^\nu +d_{2g} D^\nu \vD D^\mu) \Bigr\} \Avperpl{\nu},\\
 \L{V}{(3)} &=& (\Avperpl{\mu})^\dagger 
  \left(g^{\mu\nu}
    \left\{d_{3a} D^4 +d_{3b} \left[D^2 (\vD)^2 +(\vD)^2 D^2 \right] 
     +d_{3c} (\vD)^4 \right.\right. \nonumber \\
  &&+d_{3d} (D_\alpha \vD D^\alpha \vD +\vD D_\alpha \vD D^\alpha)
    +d_{3e} D_\alpha (\vD)^2 D^\alpha +d_{3f} D_\alpha D^2 D^\alpha \nonumber\\
  &&+\left. d_{3g} D_\alpha D_\beta D^\alpha D^\beta \right\} 
    +d_{3h} (D^\mu D^\nu D^2 +D^2 D^\mu D^\nu)
    +d_{3i} (D^\nu D^\mu D^2 +D^2 D^\nu D^\mu) \nonumber\\
  &&+d_{3j} D^\mu D^2 D^\nu +d_{3k} D^\nu D^2 D^\mu
    +d_{3l} (D^\mu D_\alpha D^\nu D^\alpha +D^\alpha D^\mu D_\alpha D^\nu)
   \nonumber\\
  &&+d_{3m} D_\alpha D^\mu D^\nu D^\alpha
    +d_{3n} D^\alpha D^\nu D^\mu D_\alpha 
    +d_{3o} (D_\alpha D^\nu D^\alpha D^\mu +D^\nu D^\alpha D^\mu D_\alpha)
     \nonumber\\
  &&+d_{3p} \left[D^\mu D^\nu (\vD)^2 +(\vD)^2 D^\mu D^\nu \right]
    +d_{3q} \left[D^\nu D^\mu (\vD)^2 +(\vD)^2 D^\nu D^\mu \right] \nonumber\\
  &&+d_{3r} D^\mu (\vD)^2 D^\nu +d_{3s} D^\nu (\vD)^2 D^\mu
    +d_{3t} \vD D^\mu D^\nu \vD \nonumber\\
  &&+d_{3u} \vD D^\nu D^\mu \vD \Bigr)
  \Avperpl{\nu}, \\
 \L{V}{(4)} &=& i(\Avperpl{\mu})^\dagger 
  \left(g^{\mu\nu}
    \left\{ d_{4a} (D^4 \vD +\vD D^4) +d_{4b} D^2 \vD D^2 \right.\right.
     \nonumber\\
  && +d_{4c} \left(D^2 D^\a \vD D_\a +D^\a \vD D_\a D^2 \right)
     +d_{4d} \left(D_\a D^2 \vD D^\a +D^\a \vD D^2 D_\a \right)\nonumber\\
  && +d_{4e} D_\a D_\b \vD D^\a D^\b +d_{4f} D_\a D_\b \vD D^\b D^\a
      \nonumber \\
  && +d_{4g} (D_\a D_\b D^\a \vD D^\b +D^\b \vD D^\a D_\b D_\a)
     +d_{4h} \left[D^2 (\vD)^3 +(\vD)^3 D^2 \right] \nonumber\\
  && +d_{4i} \, D_\a (\vD)^3 D^\a +d_{4j} (\vD)D_\a (\vD) D^\a (\vD)\nonumber\\
  && +d_{4k} \left[\vD D^2 (\vD)^2 +(\vD)^2 D^2 \vD \right]
     +d_{4l} (\vD)^5 \Bigr\} \nonumber\\
  && \left.+\hbox{constraint-type operators} \right)
  \Avperpl{\nu}. 
\endeqal{LgenHVEFT}
\reseteq
Constraint-type operators are those where covariant derivatives are 
contracted with the vectors fields. 

Now requiring that the theory be unchanged by the reparameterization in 
eq.~(\ref{vVR}-\ref{kVR}) gives the following relations
\alpheq
\eqa
 d_{1a} = 1, \nonumber\\
 d_{1b} - d_{2a} - 2d_{2b} = -1, \nonumber\\
 d_{2b} + 2d_{3a} + d_{3f} + d_{3g} = -\frac{1}{2}, \nonumber\\
 \frac{d_{1b}}{2} + d_{2c} + 2d_{3b} + d_{3d} + d_{3e} = 0, \nonumber\\
 d_{2a} + 2d_{3f} + 2d_{3g} = 0, \nonumber\\
 d_{2c} + 2d_{3d} = 0, \nonumber\\
 d_{3b} - 2d_{4a} - d_{4c} = {1 \over 4}, \nonumber\\
 d_{3c} - d_{4j} - 2d_{4k} = -{1 \over 4}, \label{HVEFTRIpred}\\
 -{d_{2a}\over 2} + d_{3d} - 2d_{4c} - d_{4e} - d_{4f} - d_{4g} = 0,\nonumber\\
 -{d_{2b}\over 2} + d_{3b} - 2d_{4b} - d_{4c} - d_{4d} - d_{4g} = 0,\nonumber\\
 {d_{2b} \over 2} + 2d_{4a} + d_{4d} = 0, \nonumber\\
 -{d_{2c} \over 2} + d_{3c} - 2d_{4h} - d_{4i} = 0, \nonumber\\
 d_{3d} = d_{4g}, \nonumber\\
 d_{3e} - 2d_{4d} - d_{4e} - d_{4f} - d_{4g} = 0. \nonumber
\endeqa
There are also relations between the constraint type terms above; for example,
\eq d_{1c} = -d_{1d}. \endeql{HVEFTRIpredconstr}
\reseteq

For consistency it should verified that the tree-level matching performed 
in the functional integral formalism of ref. \cite{Lee97a} obey these 
constraints. From that calculation, one finds
\eqa
 d_{1a}= -d_{1b}= d_{1c}= -d_{1d}= d_{2d}= d_{3j}= -d_{3r}= -d_{3t}= 1,
  \nonumber\\
 -d_{3a}= d_{3b}= -d_{3c}= -d_{3h}= d_{3i}=d_{3p}=d_{3q}={1\over 4},\nonumber\\
 d_{4b} = -d_{4h} = d_{4l} = {1\over 8}, \nonumber\\
 \hbox{and for }\rho =\hbox{2a-2c, 2e, 2f, 2g, 3d-3i, 3k-3q, 3s, 3u, 4a,
  4c-4g, 4i-4k, } d_\rho =0,
\endeqa
which satisfy the conditions in \eqr{HVEFTRIpred} and 
(\ref{HVEFTRIpredconstr}) as they should. 

The results presented here seem to be different from those in ref. 
\cite{LuM92}. In that paper, Luke and Manohar used Lorentz transformation 
properties to arrive at an effective theory field which simply picks up a 
phase under reparameterization. Hence they were able to write down the 
most general reparameterization invariant Lagrangian symbolically in closed 
form expressed in terms of this field. In the formulation given above, 
the transformation of the effective theory field $\Avperpl{\mu}$ or 
$\Avpl{\mu}$ is given by eq. (\ref{AvplVRexact}-\ref{AvplVR}) and there is 
no field constructed from combinations of $\Avperpl{\mu}$ which transforms 
by only a phase. Thus it is not possible here to achieve what they did 
except to follow the approach given above.

\section{Summary}

The velocity reparameterization invariance of heavy particles effective 
field theories is a consequence of the observation that only the total 
momentum is unique and not in how it is split into two pieces as in 
\eqr{kdef}. The implications of this invariance has been examined for 
particles of different spin. It places unusually strong constraints on 
such theories by relating the short-distance coefficients of operators of 
different dimension. Although these coefficients can be calculated by 
other means such as (perturbative) matching with renormalization group 
running, it is important to note, however, that while the values obtained 
must be determined order-by-order in the loop expansion, the predictions of 
VRI are exact results which hold to arbitrary order in perturbation theory.

\vskip 1cm 
\leftline{\large\bf Acknowledgement} \bigskip
This work was supported in part by NSERC and the Department of Energy under 
contract DOE-FG03-90ER40546.

\end{document}